\documentclass[12pt,a4j]{article}
\usepackage[dvips]{graphicx}
\usepackage{amsmath}
\usepackage{enumerate}
\begin{document}
\baselineskip=7mm
\centerline{\bf Periodic solutions of a resistive model for nonlocal
Josephson dynamics}\par
\bigskip
\centerline{Yoshimasa Matsuno}\par
\centerline{\it Division of Applied Mathematical Science, Graduate School of Science and Engineering} \par
\centerline{\it Yamaguchi University, Ube 755-8611, Japan} \par
\bigskip
\centerline{E-mail: matsuno@yamaguchi-u.ac.jp}\par
\bigskip
\noindent{\bf  Abstract}\par
\noindent A novel method is developed for constructing periodic solutions of a model equation
describing nonlocal Josephson electrodynamics. This method consists of reducing
the equation to a system of linear ordinary differential equations through a
sequence of nonlinear transformations. The periodic solutions are then obtained by
a standard procedure which are represented in terms of trigonometric functions.
It is found that the large time asymptotic of the solution exhibits a steady
profile which does not depend on initial conditions.
  \par
\bigskip
\bigskip
\leftline{\bf  1. Introduction} \par
\noindent The recent studies on Josephson tunnel junctions with high-temperature superconductors
reveal that the nonlocal nature of Josephson electrodynamics becomes dominant when the
Josephson penetration depth $\lambda_J$ is shorter than the London penetration depth $\lambda_L$.
In particular, if we consider a thin layer between two superconductors, the phase difference
$\phi(x,t)$ across the Josephson junction is described by the following model equation [1-5]
$$\omega_J^{-2}\phi_{tt}+\omega_J^{-2}\eta \phi_t=-\sin\,\phi
+{\lambda_J^2\over \pi\lambda_L}\int_{-\infty}^\infty K_0\left({|x-x^\prime|\over \lambda_L}\right)
\phi_{x^\prime x^\prime}(x^\prime,t)dx^\prime+\gamma. \eqno(1)$$
Here, $K_0$ is the modified Bessel function of order zero, $\omega_J$ is the Josephson plasma frequency, $\eta$ is
a positive parameter inversely proportional to the resistance of a unit area of the tunneling
junction, $\gamma$ is a bias current density across the junction normalized by the Josephson critical current density and the
subscripts $t$ and $x^\prime$ appended to $\phi$ denote partial differentiation.
When the characteristic space scale $l$ of $\phi$ is extremely large compared with $\lambda_L$, the
kernel $K_0$ has an approximate expression $K_0(x)\sim \pi\delta(x)$ where $\delta(x)$ is Dirac's delta function.
Then, the equation (1) reduces to the perturbed sine-Gordon equation [6].  In the opposite limit
$l<<\lambda_L$, one can use the asymptotic of the kernel $K_0(|x|)\sim -{\rm ln}\,|x|$.
In addition, if we restrict our consideration to the overdamped case $\eta>>1$ as well as
the zero bias current $\gamma=0$, then unlike the perturbed sine-Gordon equation, the equation (1) becomes
an integrodifferential (or nonlocal) equation. It can be written in an appropriate dimensionless form as
$$\phi_t=-\sin\,\phi+H\phi_x, \qquad H\phi_x={1\over \pi}P\int_{-\infty}^\infty
{\phi_{x^{\prime}}(x^{\prime},t)\over x^{\prime}-x}dx^{\prime}, \eqno(2)$$
where $H$ is the Hilbert transform operator. Equation (2) may be termed a resistive model
for nonlocal Josephson electrodynamics [7]. Note that the equation (2) has been proposed for the first
time in searching integrable nonlinear equations with dissipation [8]. The general multikink solutions of
the equation (2) have been obtained and their properties have been investigated in detail [8]. \par
 In this paper, we report some new results concerning periodic solutions of the equation (2).
    Specifically, we show that the equation (2) can be transformed to a finite-dimensional 
   nonlinear dynamical system through a dependent variable transformation. We then linearize the
   system of equations to derive a first-order system of {\it linear} ordinary
   differential equations (ODEs). Its initial value problem can be solved explicitly to obtain
   periodic solutions. It is shown that the large time asymototic of 
   the periodic solution relaxes to a steady profile independent of
   initial conditions. \par
   \bigskip
\leftline{\bf 2. Exact method of solution}\par
\leftline{\it 2.1. A nonlinear dynamical system}\par
\noindent We seek periodic solution of (2) of the form
$$\phi=i\,{\rm ln}\,{f^*\over f}, \qquad
f=\prod_{j=1}^N{1\over \beta}\sin\,\beta(x-x_j), \eqno(3)$$
where $x_j=x_j(t)$ are complex functions of $t$ whose imaginary parts are all positive, $\beta$ is a positive
parameter, $N$ is an arbitrary positive integer and $f^*$ denotes the complex conjugate expression of $f$. Using a formula for the Hilbert
transform, one has $H\phi_x=-({\rm ln}\,f^*f)_x$. Substitution of this expression and (3) into (2)
gives the following bilinear equation for $f$ and $f^*$:
$$i(f^*_tf-f^*f_t)={i\over 2}(f^2-{f^*}^2)-f_x^*f-f^*f_x. \eqno(4)$$
We divide (4) by $f^*f$, substitute $f$ from (3) and then evaluate the residue at $x=x_j$ on both sides
to obtain a system of nonlinear ODEs for $x_j$
$$\dot{x}_j=-{1\over 2\beta}{\prod_{l=1}^N\sin\,\beta(x_j-x_l^*)
\over\prod_{\substack{l=1\\(l\not=j)}}^N\sin\,\beta(x_j-x_l)}+i,\qquad j=1, 2, ..., N, \eqno(5)$$
where an overdot denotes differentiation with respect to $t$. Note that a dynamical system
corresponding to the multikink solution is derived simply from (5) by taking the limit
$\beta\rightarrow 0$ [8]. As in the multikink case, it will be demonstrated from (5) that the imaginary part of $x_j$
remains positive if it is positive at an initial time.\par
 Before proceeding, it is convenient to introduce
some notations:
$$z=e^{2i\beta x},\qquad \xi_j=e^{2i\beta x_j},\qquad \eta_j=e^{2i\beta x_j^*}, \qquad j=1, 2, ..., N, \eqno(6)$$
$$s_1=\sum_{j=1}^Nx_j,\qquad s_2=\sum_{j<l}^Nx_jx_l,\quad ...,\quad s_N=\prod_{j=1}^Nx_j, \eqno(7)$$
$$u_1=\sum_{j=1}^N\xi_j,\qquad u_2=\sum_{j<l}^N\xi_j\xi_l,\quad ...,\quad u_N=\prod_{j=1}^N\xi_j, \eqno(8)$$
$$v_1=\sum_{j=1}^N\eta_j,\qquad v_2=\sum_{j<l}^N\eta_j\eta_l,\quad ...,\quad \ v_N=\prod_{j=1}^N\eta_j, \eqno(9)$$
$$t_j=\sum_{l=1}^N\xi_l^j,\qquad j=1, 2, ..., N. \eqno(10)$$
Here, $s_j, u_j$ and $v_j$ are elementary symmetric functions of $x_l, \xi_l$ and $\eta_l\ (l=1, 2, ..., N)$, respectively.
In terms of $u_j (j=1, 2, ..., N)$ and $s_1$, $f$ from (3) can be written as
$$f={e^{-i\beta(Nx-s_1)}\over (2\beta i)^N}\left(z^N-u_1z^{N-1}+u_2z^{N-2}+ ... + (-1)^Nu_N\right). \eqno(11)$$
Thus, $u_j\ (j=1, 2, ..., N)$ and $s_1$ determine the function $f$ completely. In the following analysis,
we derive a system of equations for $u_j$. To this end, we find it appropriate to rewrite (5) in terms of $\xi_j$
and $\eta_j$ as
$$\dot{\xi}_j=-{1\over 2}\alpha u_N{\prod_{l=1}^N(\xi_j-\eta_l)
\over\prod_{\substack{l=1\\(l\not=j)}}^N(\xi_j-\xi_l)}-2\beta\xi_j,\qquad \qquad j=1, 2, ..., N, \eqno(12)$$
where
$$\alpha=\prod_{j=1}^N(\xi_j\eta_j)^{-1/2}=e^{-i\beta(s_1+s_1^*)},
\qquad u_N=\prod_{j=1}^N\xi_j=e^{2i\beta s_1}. \eqno(13)$$
Later, it will be shown that $\alpha$ is a constant independent of $t$ and $u_N$ obeys a single nonlinear ODE. \par
\bigskip
\leftline{\it 2.2. Linearization} \par
\noindent Here, we show that the  system of nonlinear ODEs (12) can be linearized in terms of
the variables $u_j$  defined by (8). We multiply $\xi_j^{n-1}$ on both sides of (12) and sum up
with respect to $j$ from $1$ to $N$ to obtain
$${1\over n}\dot{t}_n=-{\alpha\over 2}u_N\sum_{s=0}^n(-1)^sv_sI_{n-s}-2\beta t_n,
\qquad n=1, 2, ..., N, \eqno(14)$$
where $I_{n-s}$ is defined by
$$I_{n-s}=\sum_{j=1}^N\ {\xi_j^{N+n-s-1}\over\prod_{\substack{l=1\\(l\not=j)}}^N(\xi_j-\xi_l)}. \eqno(15)$$
In deriving (14), we have used the identity
$$I_n=0,\qquad -N+1\leq n\leq -1. \eqno(16)$$ \par
The time evolution of $u_n$ follows from (14) with the help of the formulas [9]
$$u_n={(-1)^{n-1}\over n}\sum_{j=0}^{n-1}(-1)^ju_jt_{n-j}, \quad 1\leq n\leq N, 
\qquad \sum_{j=0}^n(-1)^ju_jI_{n-j}=0, \quad n\geq 1, \eqno(17)$$
where $u_0=1$ and $I_0=1$. In fact, differentiating the first formula in (17) by $t$ and substituting (14) for $\dot{t}_{n-j}$,
we can show that the quantity $h_n$ defined by
$$h_n=\dot{u}_n+{\alpha\over 2}u_Nu_n-{\alpha^{-1}\over 2}u_{N-n}^*+2\beta nu_n,
\quad n=1, 2, ..., N, \eqno(18)$$
 satisfies the relation
 $$h_n={(-1)^{n-1}\over n}\sum_{j=0}^{n-1}(-1)^jh_jt_{n-j}+{(-1)^{n+1}r_n\over 2n\alpha }, \eqno(19)$$
 where
 $$r_n=\sum_{j=1}^nu_{N-j+n}^*\left[-\sum_{s=1}^{j}(-1)^{n-s}sI_{j-s}+(-1)^{n-j}t_j\right]. \eqno(20)$$
 A straightforward calculation using (17) shows that the quantity in the brackets on the
 right-hand side of (20) vanishes identically so that $r_n\equiv 0$. It follows from this and (19) that
 $$h_n={(-1)^{n-1}\over n}\sum_{j=0}^{n-1}(-1)^jh_jt_{n-j}, \qquad n=1, 2, ..., N. \eqno(21)$$
Note from (13) and (18) that $h_0=\alpha u_N/2-u_N^*/(2\alpha)=0$ which, combined with (21), leads to
the relations $h_n\equiv 0\ (n=1, 2, ..., N)$. Thus, we see that $u_n$ evolves according to the
following system of ODEs
$$\dot{u}_n+{\alpha\over 2}u_Nu_n-{\alpha^{-1}\over 2}u_{N-n}^*+2\beta nu_n=0, 
\qquad n=1, 2, ..., N. \eqno(22)$$
It is remarkable that $u_N$ obeys a single nonlinear ODE of the form
$$\dot{u}_N+{\alpha\over 2}u_N^2-{\alpha^{-1}\over 2}+2\beta Nu_N=0,  \eqno(23)$$
and other $N-1$ variables $u_1, u_2, ..., u_{N-1}$ constitute a system of
{\it linear} ODEs.
Substituting (13) into (23), we can put (23) into a nonlinear ODE for $s_1$
$$\dot{s}_1={1\over 2i\beta}\sinh(2\beta\, {\rm Im}\, s_1)+iN, \eqno(24)$$
where ${\rm Im}\, s_1$ implies the imaginary part of $s_1$. \par
\bigskip
\leftline{\bf 3. Periodic solutions}\par
\noindent The first step for constructing periodic solutions is to integrate (24).
It follows from the real and imaginary parts of (24) that
$${\rm Re}\,\dot{s}_1=0, \qquad {\rm Im}\,\dot{s_1}=-{1\over 2\beta}\sinh(2\beta\,{\rm Im}\,s_1)+N. \eqno(25)$$
Thus, the real part of $s_1$ becomes a constant ${\rm Re}\, s_1(t)={\rm Re}\, s_1(0)\equiv b$  whereas integration of the
equation for ${\rm Im}\,s_1$ yields an explicit expression. In terms of a new variable $y=2\beta\, {\rm Im}\, s_1$,
it is given by
$$e^{-y}={2\nu_N\left(-\tanh\,{y_0\over 2}+1\right)\cosh\,{\nu_Nt}
+\left\{(2\beta N+1)\tanh\,{y_0\over 2}-2\beta N+1\right\}\sinh\,{\nu_Nt}
\over
2\nu_N\left(\tanh\,{y_0\over 2}+1\right)\cosh\,{\nu_Nt}
+\left\{(2\beta N-1)\tanh\,{y_0\over 2}+2\beta N+1\right\}\sinh\,{\nu_Nt}}, \eqno(26)$$
where $\nu_N=\sqrt{(\beta N)^2+(1/4)}$ and $y_0=y(0)=2\beta\, {\rm Im}\, s_1(0)$,
For $n=1, 2, ..., N-1$, on the other hand, (22) can be written in the form
$$\dot{u}_n=-\left({1\over 2}e^{-2\beta {\rm Im}\,s_1}+2\beta n\right)u_n+{\alpha^{-1}\over 2}u_{N-n}^*. \eqno(27)$$
Note from (13) and ${\rm Re}\, s_1=b$ that $\alpha=e^{-2i\beta b}$ becomes a constant. 
The solution of the initial value problem for (27) can be obtained by means of a
standard procedure. It can be put into the form of a rational function 
$$u_n(t)={G_n\over F}, \qquad n=1, 2, ..., N-1, \eqno(28)$$
with
$$F=2\nu_N\left(\tanh\,{y_0\over 2}+1\right)\cosh\,{\nu_Nt}
+\left\{(2\beta N-1)\tanh\,{y_0\over 2}+2\beta N+1\right\}\sinh\,{\nu_Nt}, \eqno(29)$$
$$G_n=2\nu_N\left(\tanh\,{y_0\over 2}+1\right)\biggl[u_n(0)\cosh\,{\nu_nt}
+{1\over \nu_n}\left\{\beta(N-2n)u_n(0)+{\alpha^{-1}\over 2}u_{N-n}^*(0)\right\}
\sinh\,{\nu_nt}\biggr], \eqno(30)$$
where $\nu_n=\sqrt{\beta^2(N-2n)^2+(1/4)}$.
We see that the expression (28) with $n=N$ produces (26) and hence it can be used for all $u_n$. \par
A novel feature of the solution given above will become apparent if one explores the large time
asymptotic of the solution. Actually, it is easy to see from (28), (29) and (30) that as $t$ tends to infinity, $u_n\ (n=1, 2, ..., N)$ approach
the following limiting values
$$ u_n \rightarrow 0, \quad n=1, 2, ..., N-1, \qquad u_N \rightarrow e^{2i\beta b}(\sqrt{4(\beta N)^2+1}-2\beta N). \eqno(31)$$
The asymptotic form of $\phi$ follows from (3), (11) and (31), giving rise to
$$\phi \sim 2\,\tan^{-1}\left[{\sqrt{4(\beta N)^2+1}-1\over 2\beta N}\tan\,\beta\left(Nx-b-{N\pi\over 2\beta}\right)\right]. \eqno(32)$$
If we introduce a new variable $u$ by $u=\phi_x$, then in the limit $t\rightarrow \infty$, $u$ behaves like
$$u \sim {4(\beta N)^2\over \sqrt{4(\beta N)^2+1}+(-1)^N\cos\,2\beta(Nx-b)}. \eqno(33)$$
It is remarkable that the asymptotic form of $u$ does not depend on initial conditions except for a phase constant $b$.
It represents a train of nonlinear periodic waves with an equal amplitude. 
Since $u_1(0)\not=0$, the initial profile of $u$ has a spatial period $\pi/\beta$ whereas that corresponding to (33) is given
by $\pi/N\beta$. Therefore, as time evolves, there appear $N$ identical waves in the space interval $\pi/\beta$. 
The maximum  and minimum values of each wave are given respectively by
 $u_{max}=\sqrt{4(\beta N)^2+1}+1$ and 
$u_{min}=\sqrt{4(\beta N)^2+1}-1$. If we define the amplitude of the wave by $A=u_{max}-u_{min}$, then $A=2$,
indicating that the amplitude becomes a constant independent of the wavenumber. 
Figure 1 shows a typical time evolution
of $u$ for $N=2$ where the parameters are chosen as $\beta=0.2, u_1(0)=e^{-1.6}+e^{-0.8}, u_2(0)=e^{-2.4},\ (x_1(0)=4i, x_2(0)=2i)$. 
In this example, the wavelength of the periodic wave is 7.85. 
As expected from the asymptotic form (33), we can observe two identical waves with an amplitude 2  in the space interval 15.7
at a final stage of the time evolution. 
\par
\begin{center}
\includegraphics[width=10cm]{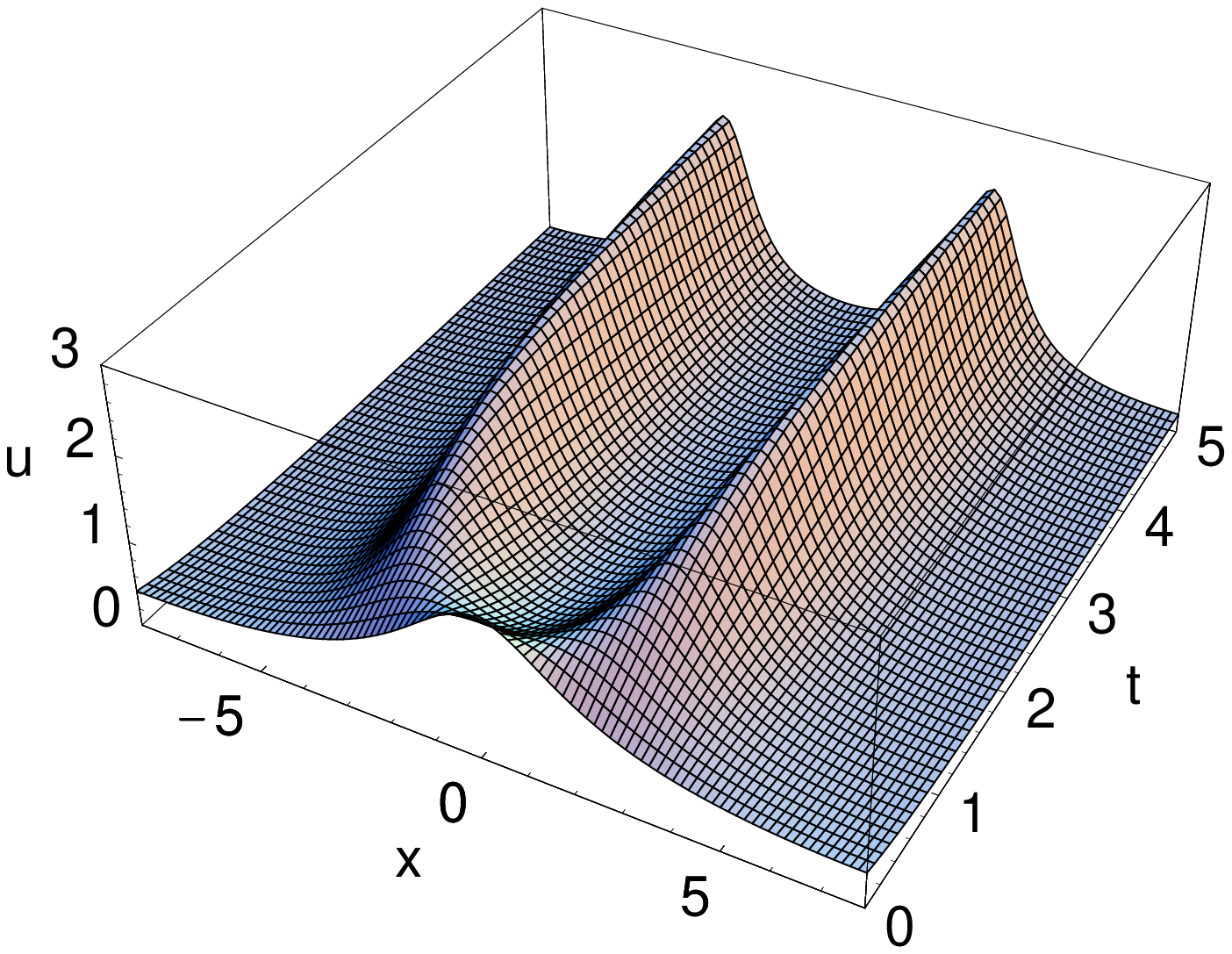}
\end{center}
\centerline{{\bf Figure 1}. Time evolution of $u$ for $N=2$ }\par
\bigskip
\leftline{\bf 4. Conclusion} \par
\noindent We have developed an exact method for constructing periodic solutions of a resistive model
equation for nonlocal Josephson electrodynamics. Although the basic equation (2) is a highly nonlinear equation with
a nonlocal term, it can be linearized through a sequence  of dependent variable transformations.
We have obtained various new results. Among them, a system of  equations (22) is a crucial consequence 
for the purpose of  determining the function $f$ given by (11) in  closed form.  The large time
asymptotic of the solution exhibits a novel feature. In particular, it relaxes to a steady profile whose functional
form does not depend on initial conditions except for a phase constant. \par
We conclude this paper with some comments.
First, we point out that for $N=1$, the periodic solution presented here reproduces an existing solution which 
has been obtained by a different method [10]. The periodic solutions for $N\geq 2$ appear here for the first time.
It is interesting to see that the large time asymptotic (32) for $\phi$ satisfies the static version of (2), i.e., $H\phi_x=\sin\,\phi$.
As already pointed out [8], this equation is a model of dislocation derived by Peierls [11, 12].
Second, we can perform a similar analysis for a resistive model with a bias current $\gamma$. 
It is a relatively easy task to obtain multikink solutions following a procedure
developed in [8]. Nevertheless, the construction of periodic solutions
deserves further study. Third, the method of solution developed here is applicable to other nonlocal nonlinear
evolution equations as well. For example, we will be able to obtain periodic solutions of the sine-Hilbert
equation $H\theta_t=-\sin\,\theta, \ \theta=\theta(x,t)$.
It is noteworthy that the construction of periodic solutions of the sine-Hilbert equation 
has been done by a different method, but the explicit solutions
have been presented only for $N=1$ [13]. On the other hand, our method will enable us to obtain periodic solutions
for general $N$. Fourth, the periodic solutions can be used to calculate various physical quantities such as the
current density and the electric and magnetic fields in a Josephson junction. These quantities can be compared with
experimental results for high-temperature Josephson junctions. The solutions to the various problems mentioned above
will be reported in a subsequent paper as well as a detailed description of the present short communication. \par
\bigskip
\newpage
\leftline{\bf References}\par
\begin{enumerate}[{[1]}]
\item Aliev Yu M, Silin V P and Uryupin S A 1992 Superconductivity {\bf 5} 230-6
\item Gurevich A 1992 Phys. Rev. {B \bf 46} 3187-90
\item Aliev Yu M and Silin V P 1993 JETP {\bf 77} 142-7
\item Gurevich A 1993 Phys. Rev. {B \bf 48} 12857-65
\item Mints R G 1997 J. Low Tem. Phys. {\bf 106} 183-92
\item Barone A and Paterno G 1982 {\it Physics and Application of the Josephson Effect} (New York: Wiley)
\item Silin V P and Uryupin S A 1995 JETP {\bf 81} 1179-91
\item Matsuno Y 1992 J. Math. Phys. {\bf 33} 3039-45
\item Matsuno Y 2004 J. Math. Phys. {\bf 45} 795-802
\item Alfimov G L and Silin V P 1994 JETP {\bf 79} 369-76
\item Peierls R 1940 Proc. Phys. Soc.  {\bf 52} 34-7
\item Nabarro F R N 1947 Proc. Phys. Soc.  {\bf 59} 256-72
\item Matsuno Y 1987 Phys. Lett. {\bf A 120} 187-90

\end{enumerate}

\end{document}